\documentclass[fleqn,usenatbib]{mnras}

\usepackage[T1]{fontenc}

\DeclareRobustCommand{\VAN}[3]{#2}
\let\VANthebibliography\thebibliography
\def\thebibliography{\DeclareRobustCommand{\VAN}[3]{##3}\VANthebibliography}

\usepackage{graphicx}	
\usepackage{amsmath}	
\usepackage{amssymb}	

\newcommand\mabs{M}
\newcommand\vol{V}

\newcommand\hstenvreal{\delta}
\newcommand\sigmapb{\sigma_{\textrm{PB}}}
\newcommand\sigmapbi{\sigma_{\textrm{PB,i}}}

\newcommand\epcv{\varepsilon_{\textrm{cv}}}
\newcommand\epcvi{\varepsilon_{{\textrm{cv}}, i}}

\newcommand\snum{N_{\textrm{f}}}

\newcommand\pakidge{\texttt{galcv}}

\title[HST Large-scale Structure]{A joint measurement of galaxy luminosity functions and large-scale field densities during the Epoch of Reionization}

\author[Trapp \& Furlanetto]{
A.C. Trapp$^{1}$\thanks{E-mail: atrapp@astro.ucla.edu} \&
Steven R. Furlanetto$^{1}$
\\
$^{1}$Department of Physics and Astronomy, University of California Los Angeles, CA, 90095-1562, USA \\
}

\date{Accepted XXX. Received YYY; in original form ZZZ}

\pubyear{2022}

\begin{document}
\label{firstpage}
\pagerange{\pageref{firstpage}--\pageref{lastpage}}
\maketitle

\begin{abstract}
One of the most exciting advances of the current generation of telescopes has been the detection of galaxies during the epoch of reionization, using deep fields that have pushed these instruments to their limits. It is essential to optimize our analyses of these fields in order to extract as much information as possible from them. 
In particular, standard methods of measuring the galaxy luminosity function discard information on large-scale dark matter density fluctuations, even though this large-scale structure drives galaxy formation and reionization during the Cosmic Dawn. 
Measuring these densities would provide a bedrock observable, connecting galaxy surveys to theoretical models of the reionization process and structure formation.
Here, we use existing Hubble deep field data to simultaneously fit the universal luminosity function and measure large-scale densities for each Hubble deep field at $z =$ 6--8 by directly incorporating priors on the large-scale density field and galaxy bias.
Our fit of the universal luminosity function is consistent with previous methods but differs in the details. For the first time, we measure the underlying densities of the survey fields, including the most over/under-dense Hubble fields. We show that the distribution of densities is consistent with current predictions for cosmic variance.
This analysis on just 17 fields is a small sample of what will be possible with the James Webb Space Telescope, which will measure hundreds of fields at comparable (or better) depths and at higher redshifts.

\end{abstract}




\section{Introduction}

Over the past three decades, astronomers have put enormous effort -- and time with facilities like the Hubble Space Telescope (HST) -- into observing the most distant galaxies.
As we approach the era of the James Webb Space Telescope (JWST), we expect a revolution in our understanding of the early Universe. JWST's first observational campaigns will uncover many interesting and complex phenomena in the Cosmic Dawn, but interpreting these new observations will require a solid bedrock of survey analysis. 

A key observable of the Cosmic Dawn is the galaxy luminosity function, which describes the galaxy population and its growth as a whole. Much effort has been put into its study, as its evolution in shape and normalization have important implications for the ways galaxies form and evolve \citep[see e.g.,][]{Schenker2013,McLure2013,Bouwens2015, Finkelstein2015, Bowler2015, Livermore2017, Atek2018, Oesch2018, Behroozi2019, Bouwens2021, Finkelstein2022}. These studies have pinned down the abundance of relatively bright galaxies at $z \lesssim 8$, with results largely consistent with models extrapolated from lower redshift \citep[see e.g.,][]{Tacchella2013, Mason2015, Furlanetto2017, Mirocha2017}. However, above $z \gtrsim 9$, galaxies are currently too rare to decisively measure their abundances, although the observations still provide important insights into early galaxies  \citep{Oesch2013,Oesch2015,Bouwens2015, Ishigaki2015,Mcleod2015,Mcleod2016,Bouwens2019,RobertsBorsani2022}.
These measurements have been possible thanks to several large observing campaigns across a few distinct fields: only by combining many such efforts have astronomers managed to obtain the current constraints. 

\begin{figure}
    \centering
    \includegraphics[width=0.485\textwidth]{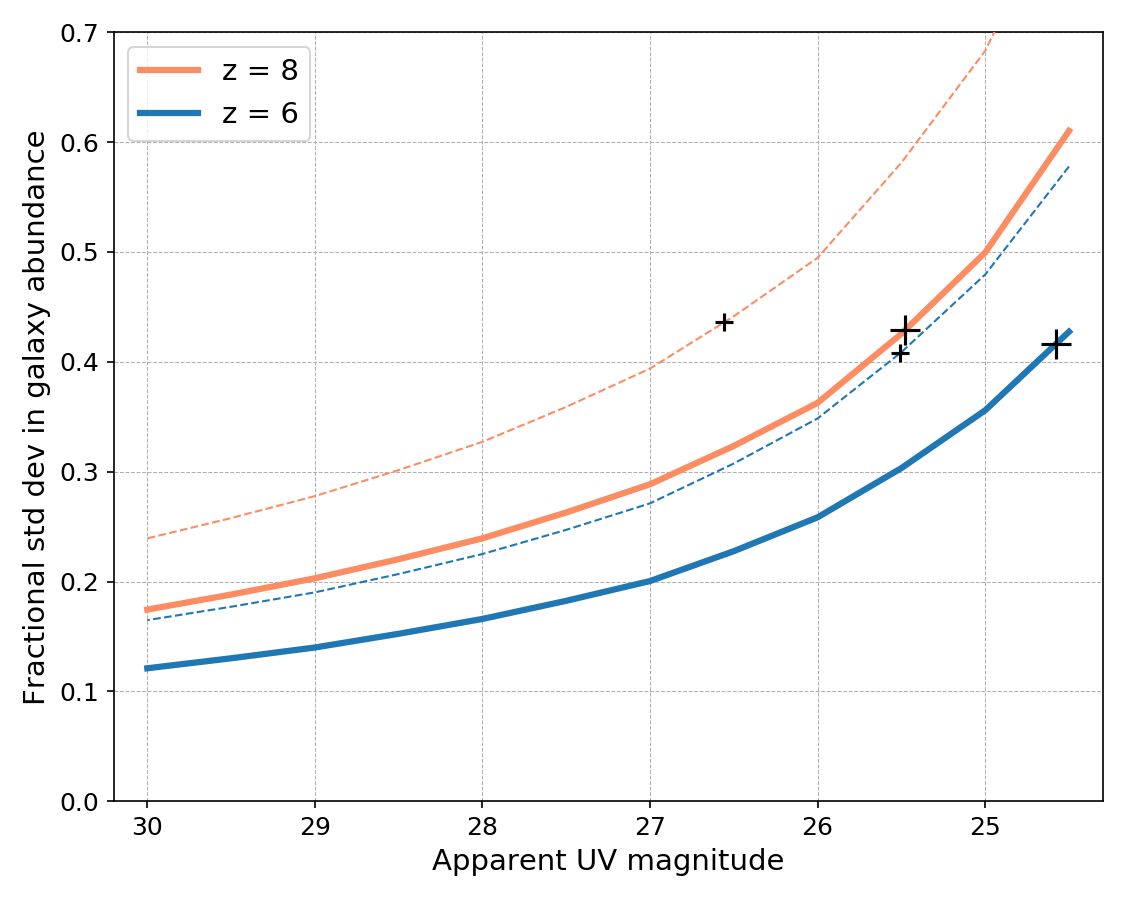}
    \caption{
    The strength of cosmic variance, or fractional standard deviation in galaxy abundance, as a function of apparent magnitude for redshifts 6 and 8. This is equivalent to the $\epcv$ parameter used in Section~\ref{hst_sec:LFmethods}.
    A value of e.g., 0.3 corresponds to a galaxy over-density of 30\% for a 1-$\sigma$ over-dense region (not accounting for Poisson noise). 
    \textit{Thick solid} lines are for a 50 arcmin$^2$ survey and \textit{thin dashed} lines are for a 5 arcmin$^2$ survey, both with $\Delta z = 1$. The strength of cosmic variance becomes more dependent on magnitude at higher redshift and for smaller volumes; cosmic variance significantly affects the shape of the luminosity function in these cases. The `+' markers show where we would expect to find $\sim $1 source at the indicated magnitude in a survey. This is to mark where Poisson noise dominates.}
    \label{hst_fig:shapechange}
\end{figure}

One of the (many) challenges in measuring the luminosity function is the uncertainty due to cosmic variance\footnote{In this paper, we use the term ``cosmic variance'' to describe dark matter density fluctuations between volumes in our Universe and the subsequent consequences for the galaxy population. To be precise, this is a case of sample variance. The term cosmic variance is sometimes reserved for the errors stemming from having only one Universe to observe.}: the normalization and shape of the luminosity function differ between distinct volumes due to fluctuations in the large-scale dark matter density field \citep[see Figure~\ref{hst_fig:shapechange}, and][]{Trapp2020}.
However, to the extent that it reflects real large-scale structure in the Universe, cosmic variance is not just a
nuisance; it is itself a key driver of both galaxy formation and reionization during the Cosmic Dawn. If large-scale densities can be measured, they can complement the luminosity function as another bedrock observable.
The insights to be gained from such measurements include:
\textit{(i)} Reionization likely began in the  densest parts of the Universe and ended in the largest voids. Identifying such over/under-densities is an area of great interest \citep[see e.g.][]{Zitrin2015,Jung2020,Tilvi2020,Hu2021,Endsley2021,Becker2018, Davies2018, Christenson2021}. 
\textit{(ii)} Large-scale feedback mechanisms, driven by large-scale structure, are likely to strongly affect the galaxy population before and during reionization \citep{Thoul1996, Iliev2007, Noh2014}. 
\textit{(iii)} Measuring large-scale densities at early times facilitates the understanding of the assembly history of rare objects like galaxy clusters (e.g., \citealt{Chiang2017}), which form from the densest environments.
\textit{(iv)} Finally, comparing large-scale density measurements from surveys with theoretical predictions of cosmic variance can help test models of the galaxy--halo connection \citep{Trapp2020}.

In \citet{Trapp2022}, we developed a framework that simultaneously measures field densities \textit{and} the high-$z$ luminosity function given a set of galaxy surveys. Unlike the standard approach to estimating luminosity functions, which acknowledges the existence of cosmic variance but does not attempt to model it, our new framework uses Bayesian statistics to fold in a comprehensive model of cosmic variance \citep{Trapp2020} and its effect on the galaxy population (which changes the shape of the luminosity function, see Figure~\ref{hst_fig:shapechange}). As a result, our method also measures the large-scale densities of the survey fields. We then predicted the precision of various JWST cycle-1 surveys, finding that these surveys can measure field densities to the maximum precision allowed by Poisson noise. We also found these surveys can measure the luminosity function at $z =$ 12 with comparable precision to HST's existing constraints at $z =$ 8, but only if the data sets can be combined effectively.

In this paper, we apply that same framework to existing HST galaxy data from \citet{Bouwens2021}\footnote{For $z =$ 6--8, the data-set from \citet{Bouwens2021} is the same as \citet{Bouwens2015} with the addition of the COSMOS, UDS, and EGS fields.} and \citet{Finkelstein2015} (see Table~\ref{hst_tab:fields}). We obtain a new measurement of the galaxy luminosity function for $z =$ 6--8 and, for the first time, measure the underlying large-scale density of every HST survey field in this data set.
This work also demonstrates the power of measuring field densities, with an eye forward to the much larger, deeper, and more comprehensive data set that will be arriving in cycle-1 of JWST, allowing for many more densities to be measured at higher precision.

After analyzing the existing data with our framework, we compare to earlier luminosity function estimates and present the environment measurements. We also develop a new method of calculating individual field densities \textit{after} a global luminosity function fit. This method drastically reduces computation time required to obtain field densities with a minimal loss in precision.

In section~\ref{hst_sec:methods}, we describe the data sets we use, briefly summarize the framework from \citet{Trapp2022}, and describe the new method of measuring environments. In sections~\ref{hst_sec:LFresults} and ~\ref{hst_sec:Envresults}, we present our new measurements of the $z =$ 6--8 luminosity function and field densities. In section~\ref{hst_sec:conclusions} we discuss our results.

We use the following cosmological parameters: $\Omega_m = 0.308$, $\Omega_\Lambda=0.692$, $\Omega_b=0.0484$, $h=0.678$, $\sigma_8=0.815$, and $n_s=0.968$, consistent with recent Planck Collaboration XIII results \citep{PlanckCollaboration2016}. We provide all distances in comoving units. We present all luminosities as rest-frame ultra-violet ($1500-2800$ \AA)\footnote{This wavelength range corresponds to $H$-band in the redshift range of $z\approx5$--$9$ and $K$-band for $z\approx8$--$12$.} luminosities, and all magnitudes are AB magnitudes.

\begin{table*}
	\centering
	\caption{The area, magnitude limit, and source counts of each field. For fields with both \citet{Bouwens2021} and \citet{Finkelstein2015} data, the latter number count is in parentheses. \citet{Finkelstein2015} uses redshift bins of size $\Delta z = 1$ centered at $z =$ 6, 7, 8. \citet{Bouwens2021} use redshift intervals of 5.5 $<z<$ 6.3 for their $z\sim$ 6 sample, 6.3 $<z<$ 7.3 for their $z\sim$ 7 sample, and 7.3 $<z<$ 8.4 for their $z\sim$ 8 sample.
	}
	\label{hst_tab:fields}
	\begin{tabular}{cccccc}
		\hline
		\hline
		Field & Area & approx depth & & Number & \\
		 & [arcmin$^2$] & [rest-UV] & z = 6 & z = 7 & z = 8 \\
		\hline
		Bouwens (Finkelstein) & & & & &\\
		& & & & &\\
		CANDELS-GS-DEEP & 64.5 & 27.9 & 198 (142) & 77 (48) & 26 (16)\\
		ERS & 40.5 & 27.8 & 61 (80) & 46 (48) & 5 (6)\\
		CANDELS-GS-WIDE & 34.2 & 27.4 & 43 (40) & 5 (4) & 3 (1)\\
		CANDELS-GN-DEEP & 68.3 & 28.1 & 188 (180) & 134 (92) & 51 (18)\\
		CANDELS-GN-WIDE & 65.4 & 27.4 & 69 (63) & 39 (24) & 18 (14)\\
		HUDF/XDF & 4.7 & 29.7 & 97 (94) & 57 (40) & 29 (15)\\
		HUDF09-1 & 4.7 & 28.8 & 38 (35) & 22 (16) & 18 (4)\\
		HUDF09-2 & 4.7 & 28.9 & 32 (31) & 23 (10) & 15 (3)\\
		MACS0416-Par & 4.9 & 29.0 & 25 (24) & 19 (8) & 4 (1)\\
		Abell 2744-Par & 4.9 & 28.9 & 20 (13) & 11 (8) & 4 (1)\\
		\hline
		Bouwens only & & & & &\\
		& & & & &\\
		CANDELS-UDS & 151.2 & 26.8 & 33 & 18 & 6\\
		CANDELS-COSMOS & 151.9 & 26.8 & 48 & 15 & 9\\
		CANDELS-EGS & 150.7 & 26.9 & 50 & 43 & 9\\
		MACS0717-Par & 4.9 & 28.8 & 41 & 21 & 10\\
		MACS1149-Par & 4.9 & 28.8 & 36 & 31 & 6\\
		Abell S1063-Par & 4.9 & 28.8 & 40 & 20 & 7\\
		Abell 370-Par & 4.9 & 28.8 & 47 & 20 & 3\\
		\hline
	\end{tabular}
\end{table*}

\section{Methods and Data}\label{hst_sec:methods}

\subsection{Fitting the luminosity function}\label{hst_sec:LFmethods}
We fit a Schechter luminosity function to galaxy catalogs using a Bayesian fitting framework. This framework is described in detail in section 2.1 and 2.2 of \citet{Trapp2022}, but we summarize it  here.

Let us assume that the average number density of galaxies with absolute magnitudes between $(\mabs, \mabs+d\mabs)$ is described by $\Phi_{\textrm{avg}}(\mabs,z) d\mabs$, which is a Schechter function with the following redshift-dependent parameters $\vec{\phi}(z)$: \textit{(i)} the normalization $\phi^*$, \textit{(ii)} the characteristic magnitude $M^*$, and \textit{(iii)} the faint-end power-law slope $\alpha$:
\begin{equation}
\begin{aligned}
    &\Phi_{\textrm{avg}}(\mabs,z) d\mabs = \\
    &(0.4~{\textrm{ln}}10)\phi^*[10^{0.4(\mabs^*-\mabs)}]^{\alpha+1}{\textrm{exp}}[-10^{0.4(\mabs^*-\mabs)}]d\mabs.
\end{aligned}
\end{equation}

The luminosity function that can actually be observed also depends on: \textit{(i)} the effects of cosmic variance and \textit{(ii)} observational features like the completeness and contamination functions (which we combine into a single function $f(\mabs, z)$) that are unique to each survey volume. The luminosity function in each survey volume becomes:
\begin{equation}\label{hst_eq:phiobs}
\begin{aligned}
    &\Phi_{\textrm{obs}}(\mabs,\vol,z,\hstenvreal)= \\ &f(\mabs, z) \cdot \Phi_{\textrm{avg}}(\mabs,z) \left( 1 + \frac{\hstenvreal}{\sigmapb}~\epcv(\mabs,\vol,z)\right).
\end{aligned}
\end{equation}
where $\hstenvreal = (\rho - \bar{\rho})/\bar{\rho}$ is the relative linear dark matter density in the volume $\vol$, $\sigmapb$ is the rms fluctuation of the linear dark matter density field on the scale $\vol$ (and `PB' refers to the pencil-beam shape typical of real surveys), and $\epcv(\mabs,\vol,z)$ parameterizes the luminosity-dependent cosmic variance using the model from \cite{Trapp2020}.

The cosmic variance function $\epcv(\mabs,V,z)$ combines non-linear halo clustering with a self-consistent analytical galaxy model (see Figure~\ref{hst_fig:shapechange} for example values of $\epcv$). It also corrects for the `pencil-beam' shape of survey volumes and is comparable to simulation-based estimates of cosmic variance from \citet{Bhowmick2020} and \citet{Ucci2021}. The largest uncertainty in $\epcv$ comes from the models of non-linear halo clustering; $\epcv$ varies $\sim$25$\%$ between models. The galaxy model can also affect $\epcv$, although to a much lesser extent because $\epcv$ is not a strong function of magnitude (at increasing redshift, however, $\epcv$'s dependence on magnitude increases). We test the impact of uncertainty in $\epcv$ on our results in section~\ref{hst_sec:measuringcosmicvariance}. 
For more information on $\epcv$, we refer the reader to \cite{Trapp2020}, where we provide a full description of the construction of $\epcv$, explore its uncertainties and model dependence, and quantitatively compare to simulation-based estimations of cosmic variance from \citet{Bhowmick2020} and \citet{Ucci2021}. We make $\epcv$ available to the public via the Python package: \pakidge.

Given data $\vec{D}$ from a large suite of galaxy surveys composed of $\snum$ fields each with their own volume, $f(\mabs, z)$, and density $\hstenvreal$, we would like to determine the probability density of the luminosity function parameters given the data: $p(\vec{\phi} | \vec{D})$, where $\vec{D}$ contains many galaxies with measured magnitudes and redshifts.. We are also interested the probability density of the dark matter densities of the $\snum$ fields given the data: $p(\vec{\hstenvreal} | \vec{D})$, where $\vec{\hstenvreal}$ is a vector containing $\hstenvreal$ for each survey.
Starting with the joint posterior $p(\vec{\phi}, \vec{\hstenvreal} | \vec{D})$ and applying Bayes' theorem, we have:
\begin{equation}
    p(\vec{\phi},\vec{\hstenvreal} | \vec{D}) \propto p(\vec{D} | \vec{\phi},\vec{\hstenvreal}) \times p(\vec{\phi}) \times p(\vec{\hstenvreal})
\end{equation}
where $p(\vec{D} | \vec{\phi},\vec{\hstenvreal})$ is the likelihood $\mathcal{L}$ given the average luminosity function parameters and densities, and $p(\vec{\phi})$ and $p(\vec{\hstenvreal})$ are their priors. We assume flat priors for each luminosity function parameter. The prior for each density $p(\hstenvreal_i)$ is simply a normal function centered at zero with standard deviation equal to $\sigmapbi$.

\begin{figure*}
    \centering
    \includegraphics[width=0.95\textwidth]{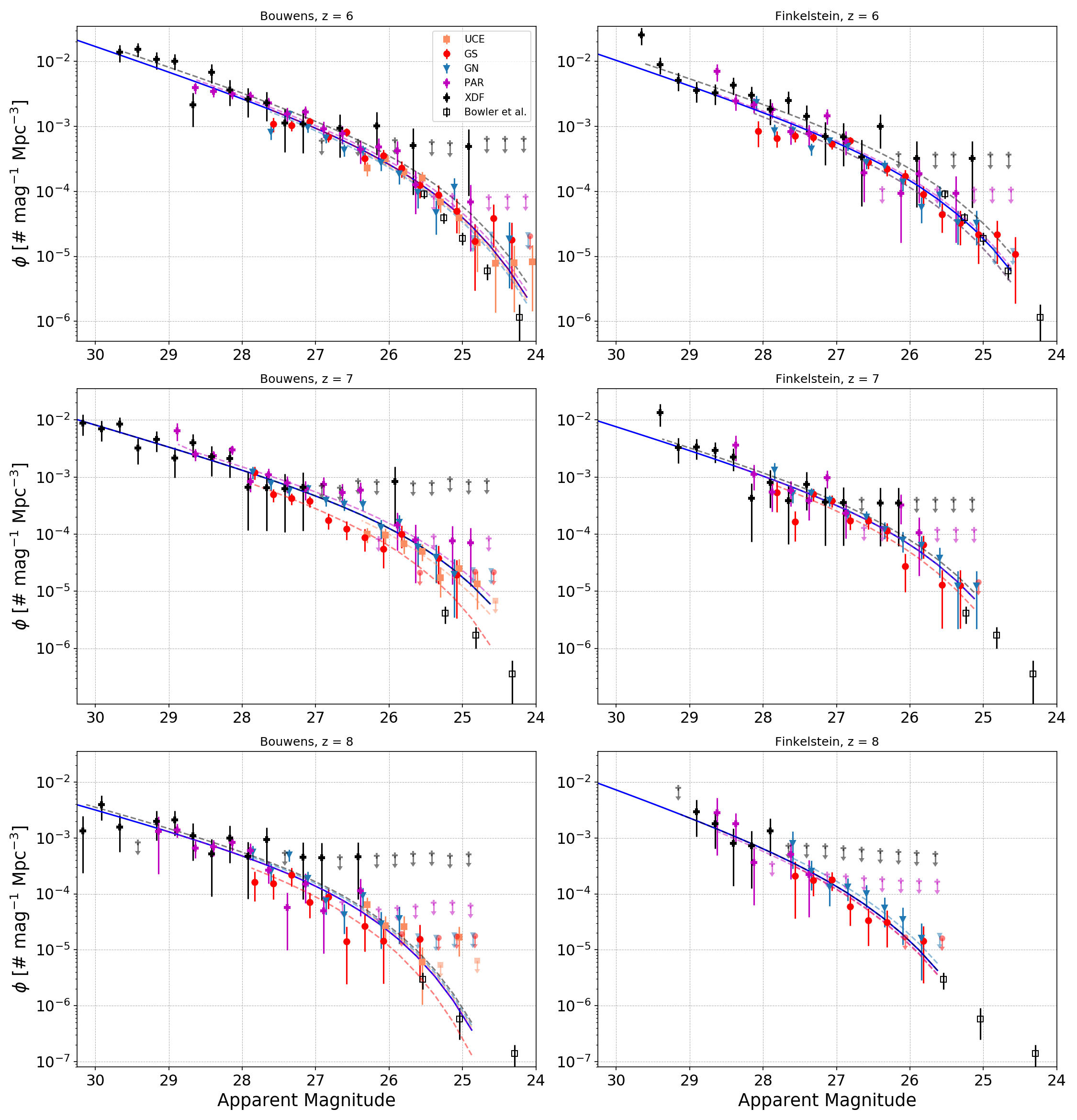}
    \caption{The best fit Schechter functions (\textit{solid blue line}) for the \citet{Bouwens2021} (\textit{left column}) and \citet{Finkelstein2015} (\textit{right column}) data sets at $z =$ 6--8 (\textit{top-to-bottom}). The dashed lines are the best-fit local luminosity functions for each composite survey. The \textit{open square} points are the binned luminosity function values from \citet{Bowler2014,Bowler2015,Bowler2020}. They lay somewhat below the best-fit lines at $z=$ 6 and especially at $z=$ 7 when compared to the \citet{Bouwens2021} data set (\textit{left column}). However, the points agree more closely with \citet{Finkelstein2015} data set (\textit{right column}). This highlights the potential effects of systematics inherent in different reduction techniques as well as space- vs ground-based measurements.
    }
    \label{hst_fig:LFs}
\end{figure*}

From \citet{Trapp2022}, the log likelihood is 
\begin{equation}\label{hst_eq:likelihood}
\begin{aligned}
    &\textrm{ln}\mathcal{L} \propto \sum_i^{\snum} \Bigg \{- n_{i,\textrm{exp}} +\\
    &\sum_{j}^{n_i} \left[\textrm{ln}\Phi_{\textrm{avg}}(M_j,\vec{\phi}) +  \textrm{ln}\left(1+\frac{\hstenvreal_{i}}{\sigmapbi}\epcvi(M_j,V_i)\right)\right] \Bigg \},
\end{aligned}
\end{equation}
where the first sum is over each field, and the second sum is over each source in the $i^{\textrm{th}}$ field. Also, $n_{i,\textrm{exp}}$ is the number of sources \textit{expected} in the $i^{\textrm{th}}$ field given the average luminosity function parameters $\vec{\phi}_{\textrm{avg}}$ and the local density $\hstenvreal_{i}$. 
We can then write the posterior as
\begin{equation}\label{hst_eq:posterior}
    p(\vec{\phi},\vec{\hstenvreal} | \vec{D}) \propto \mathcal{L} \times p(\vec{\hstenvreal}) \times p(\vec{\phi}).
\end{equation}
Finally, we can marginalize over $\vec{\phi}$ or $\vec{\hstenvreal}$ to get $p(\vec{\hstenvreal} | \vec{D})$ or $p(\vec{\phi} | \vec{D})$, respectively.

At these high redshifts, the exponential cutoff can be poorly sampled by data. If $M^*$ is brighter than the brightest galaxy in the sample, the data would be better fit by a single power-law. This results in an extreme degeneracy between the normalization $\phi^*$ and cutoff location $M^*$. To address this, we restrict the value of $M^*$ in our fits to be fainter than the brightest galaxy in our sample.

When $\epcv$ becomes large, the factor $\left( 1 + \hstenvreal / \sigmapb \cdot \epcv \right)$ can become negative for moderate under-densities, implying a negative expectation value for the number of galaxies. This is a limitation of the Gaussian approximation to cosmic variance: in reality, we must have 
$\delta \ge -1$. The dark matter fluctuations on the relevant scales (of the survey fields) are much smaller than this, but the relative density of highly-biased, luminous galaxies can reach this limit. Cosmic variance's effects on a local luminosity function are better described by a distribution function in which the probability density vanishes at negative number counts but reduces to a Gaussian when $\epcv$ is small, like a log-normal or gamma distribution. In this paper, we use the latter, and the observed luminosity function becomes
\begin{equation}\label{hst_eq:phiobsgamma}
    \Phi_{\textrm{obs}}(\mabs,\vol,z,\hstenvreal)= f(\mabs, z) \cdot f_{\Gamma}(x,k,\theta),
\end{equation}
where the gamma distribution $f_{\Gamma}$ is defined as
\begin{equation}
    f_{\Gamma}(x) = \frac{1}{\Gamma(k)\theta^k}x^{k-1}e^{-x/\theta}.
\end{equation}
The mean of this distribution is $k\theta$ and the variance is $k\theta^2$. We choose $k$ and $\theta$ such that those values match the Gaussian case: $k\theta = \Phi_{\textrm{avg}}(M,z)$ and $k\theta^2 = \Phi_{\textrm{avg}}^2(\mabs,z)\epcv^2(\mabs,\vol,z)$. $\Gamma(k)$ is the gamma \textit{function}. The variable $x$ is chosen such that the the gamma \textit{cumulative} distribution function at $x$ is equal to the normal cumulative distribution function at $\hstenvreal/\sigmapb$. This switch also carries over to the likelihood function in the following way:
\begin{equation}\label{hst_eq:likelihoodgamma}
    \textrm{ln}\mathcal{L} \propto \sum_i^{\snum} \left[- n_{i,\textrm{exp}} +
    \sum_{j}^{n_i} \textrm{ln}f_{\Gamma}(x,k,\theta) \right].
\end{equation}

\subsection{Data Sets}\label{hst_sec:data}

Our analysis makes use of existing data in the redshift range $z =$ 6--8. We use the public galaxy catalogs from \citet{Bouwens2021} and \citet{Finkelstein2015}. Table~\ref{hst_tab:fields} lists the fields used and the number counts of galaxies in those fields from each group. These catalogs contain photometric redshift and rest-frame UV magnitudes for each galaxy in their samples. We also use the completeness and contamination functions calculated 
for each field by those groups. 
These functions are obtained by simulations and become uncertain for very faint galaxies, which can affect the results. For \citet{Finkelstein2015}/\citet{Bouwens2021}, we discard all sources fainter than the magnitude at which the effective volume curve drops below 50\%/33\%, respectively.

The \citet{Bouwens2021} data set contains all the fields covered by \citet{Finkelstein2015}, plus four additional HST parallel fields and three shallow wide field surveys.
Where they overlap, these two groups start with similar raw data, but use different selection criteria and reduction pipelines.
For example, their filter thresholds select galaxies over slightly different redshift intervals, as specified in Table~\ref{hst_tab:fields}.
We consider both final data sets in parallel to compare their results and to test the robustness of our method for calculating large-scale environments in regards to these systematic choices.

A challenge of considering the density of each field is the expanded dimensionality of the parameter space; each field introduces a new density parameter to the fit.
That new parameter has a tight prior, however: a normal distribution centered at zero with standard deviation equal to $\sigmapbi$, as all of these fields are large enough to be in the linear regime. 
Unfortunately, sampling the posterior with many sub-fields can still be costly.
To alleviate this limitation, in \citet{Trapp2022} we developed a method to combine many different fields into ``composite'' fields with a single density parameter and an ``effective'' cosmic variance value, with the treatment of the combination depending on whether the fields are contiguous or treated as independent.
In this work, we combine the following groups of fields into ``composite'' fields: 
\begin{enumerate}
    \item \textbf{GS}: CANDELS-GS-DEEP, ERS, and CANDELS-GS-WIDE are combined \textit{contiguously}.    
    \item \textbf{GN}: CANDELS-GN-DEEP and CANDELS-GN-DEEP are combined \textit{contiguously}.
    \item \textbf{PAR}: HUDF09-1, HUDF09-2, MACS0416-Par, and Abell 2744-Par are combined \textit{independently}. For fits with \citet{Bouwens2021} data, this grouping also includes MACS0717-Par, MACS1149-Par, Abell S1063-Par, and Abell 370-Par.
    \item \textbf{XDF}: The HUDF/XDF field is not combined with any others.
    \item \textbf{UCE}: For fits with \citet{Bouwens2021} data, the CANDELS-UDS, CANDELS-COSMOS, and CANDELS-EGS fields are combined \textit{independently}.
\end{enumerate}
We test the effects of these combinations in section~\ref{hst_sec:validation}. In short, we find that combining fields can have a non-negligible effect, but the changes are well within the current uncertainties, so the existing HST data do not demand a more intensive treatment.

\begin{figure}
    \centering
    \includegraphics[width=0.485\textwidth]{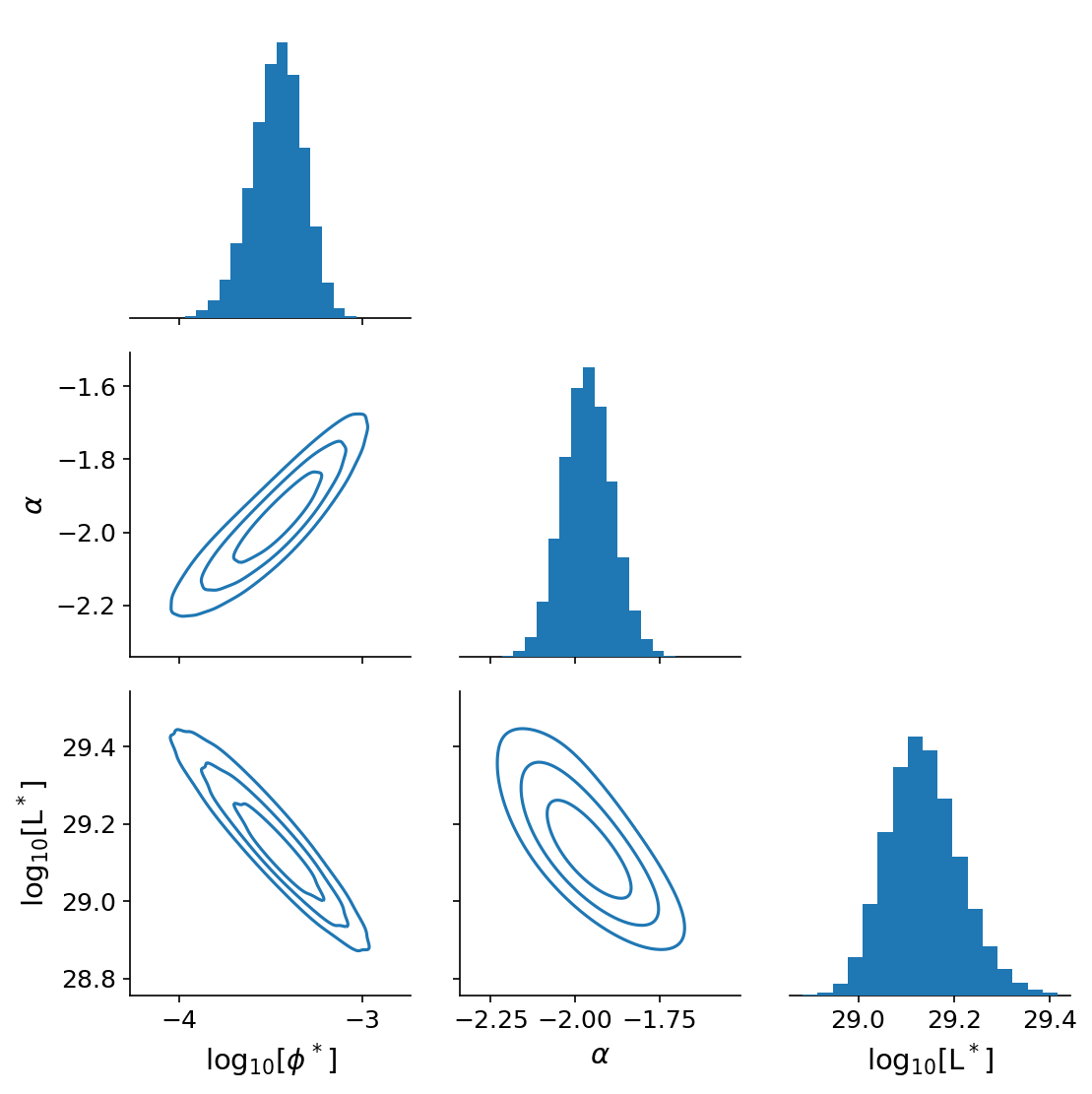}
    \caption{Our posterior of the Schechter parameters using the $z =$ 6 \citet{Bouwens2021} data set.}
    \label{hst_fig:posterior}
\end{figure}

\subsection{Measuring environments}\label{hst_sec:postprocess}

In the previous section, we combined survey fields in order to speed up the calculation of the luminosity function posterior.
Unfortunately, the densities of the individual fields are lost when combining them in this way
(except for HUDF/XDF, which is not combined with other fields).
In this section, we 
introduce a new ``post-processing'' method to measure their densities efficiently.

\begin{figure*}
    \centering
    \includegraphics[width=\textwidth]{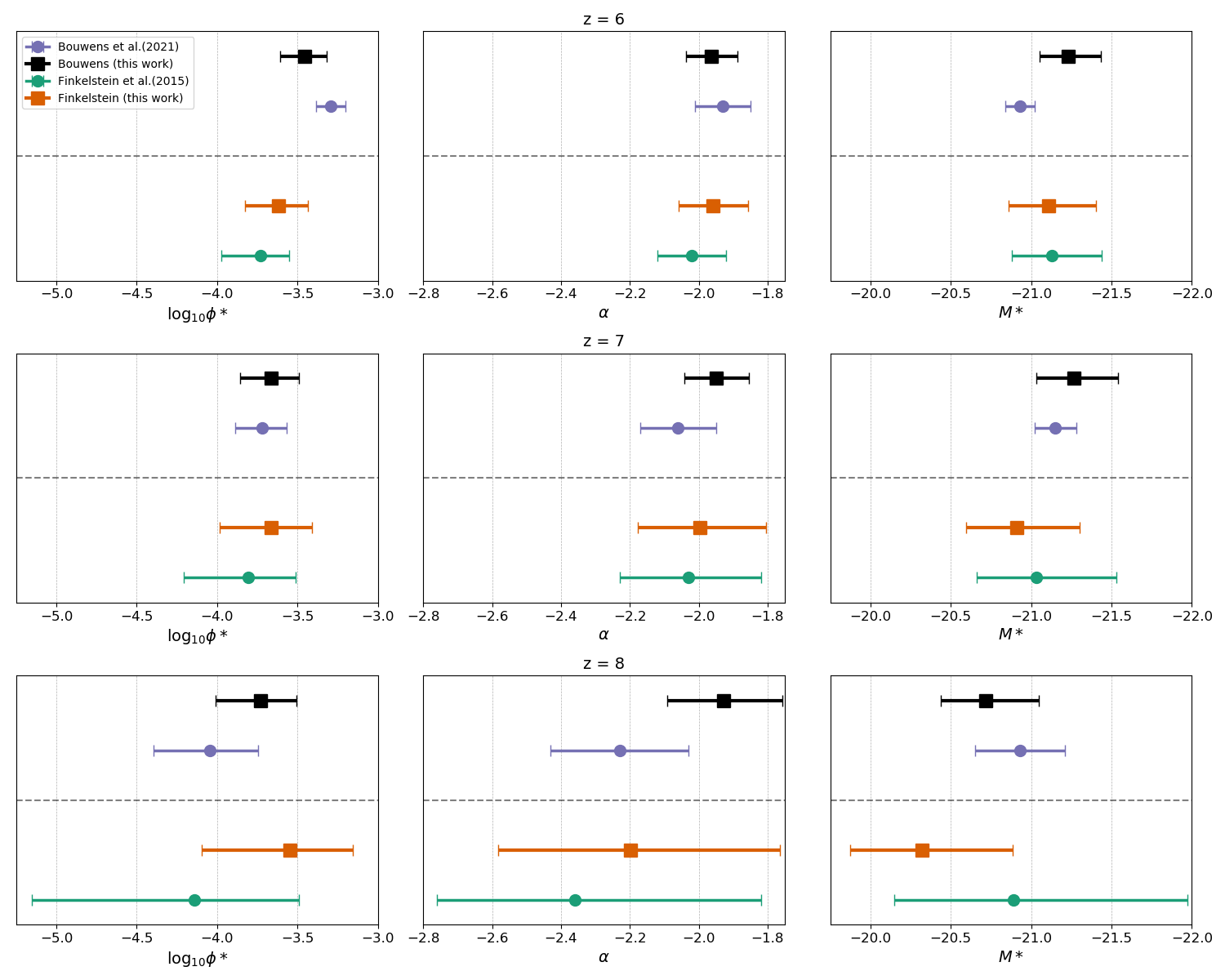}
    \caption{Marginalized posteriors of luminosity function parameters for $z =$ 6, 7, 8 (\textit{top, middle, bottom row}). The \textit{black} and \textit{orange} \textit{squares} are our results using the \citet{Bouwens2021} and \citet{Finkelstein2015} data sets, respectively.
    The \textit{purple} and \textit{green} points are the results of \citet{Bouwens2021} and \citet{Finkelstein2015}, respectively.
    These data points are also available in Table~\ref{hst_tab:params}.}
    \label{hst_fig:LFMeta}
\end{figure*}

We want $p(\hstenvreal_i | D_i)$ for the $i^{\textrm{th}}$ field. The likelihood $\mathcal{L}_i = p(D_i | \hstenvreal_i, \vec{\phi})$ of the data is similar to equation~(\ref{hst_eq:likelihood}), but applied only to one field:
\begin{equation}\label{hst_eq:likelihoodi}
    \textrm{ln}\mathcal{L}_i \propto - n_{i,\textrm{exp}} + 
    \sum_{j}^{n_i} \textrm{ln}f_{\Gamma}(x,k,\theta).
\end{equation}
The posterior for the density of the $i$th field then becomes
\begin{equation}\label{hst_eq:posteriori}
    p(\hstenvreal_i | \vec{D}) \propto p(\hstenvreal_i)\int\mathcal{L}_i \times p(\vec{\phi}) d\vec{\phi},
\end{equation}
where the prior on the Schechter parameters $p(\vec{\phi})$ is the \textit{posterior} on those parameters that were found using section~\ref{hst_sec:LFmethods}. Technically, the Schechter parameter prior $p(\vec{\phi})$ and the likelihood $\mathcal{L}_i$ are not completely independent, as the data from the $i^{\textrm{th}}$ field was used to create $p(\vec{\phi})$. The correct thing to do is recalculate the $p(\vec{\phi})$ using all fields \textit{except} the field for which we wish to measure the density. However, this procedure would require re-calculating $p(\vec{\phi})$ $\snum$ times, which would be very computationally expensive. Further, each individual field is only a small part of the data set, having a small effect on the calculation of $p(\vec{\phi})$, making them only weakly correlated.
We verify this claim in section~\ref{hst_sec:validation}.




\section{Measurements of the Luminosity Function} \label{hst_sec:LFresults}

We plot the \citet{Bouwens2021} and \citet{Finkelstein2015} data with our best-fit average luminosity functions and each composite field's luminosity function in Figure~\ref{hst_fig:LFs}. Figure~\ref{hst_fig:posterior} is an example of the Schechter function parameter posteriors produced by our framework. We provide full Schechter function fit posteriors along with this work as supplementary data (see Data Availability Section).

\begin{table*}
	\centering
	\caption{Constraints on luminosity functions from various survey combinations. All error bars are 68.27\% credible intervals}
	\label{hst_tab:params}
	\begin{tabular}{clccc}
		\hline
		\hline
		Redshift & Data Set & & Parameter Posteriors & \\
		 & & $\phi^* \times 10^4$ & $\alpha$ & $M^*$\\
		\hline
		6 & & & & \vspace{4pt}\\
		& \citet{Bouwens2021} & 5.1$_{-1.0}^{+1.2}$ & -1.93$_{-0.08}^{+0.08}$ & -20.93$_{-0.09}^{+0.09}$ \vspace{4pt}\\
		 & Bouwens & 3.5$^{+1.3}_{-1.0}$ & -1.96$^{+0.08}_{-0.07}$ & -21.23$^{+0.18}_{-0.20}$\vspace{4pt}\\
		 & \citet{Finkelstein2015} & 1.9$_{-0.8}^{+0.9}$ & -2.02$_{-0.1}^{+0.1}$ & -21.13$_{-0.31}^{+0.25}$\vspace{4pt}\\
		 & Finkelstein & 2.4$^{+1.2}_{-0.9}$ & -1.96$^{+0.10}_{-0.10}$ & -21.11$^{+0.25}_{-0.29}$ \vspace{4pt}\\
		\hline
		7 & & & &\vspace{4pt}\\
		 & \citet{Bouwens2021} & 1.9$_{-0.6}^{+0.8}$ & -2.06$_{-0.11}^{+0.11}$ & -21.15$_{-0.13}^{+0.13}$ \vspace{4pt}\\
		 & Bouwens & 2.2$^{+1.1}_{-0.8}$ & -1.95$^{+0.10}_{-0.09}$ & -21.26$^{+0.23}_{-0.28}$\vspace{4pt}\\
		 & \citet{Finkelstein2015}  & 1.6$_{-1.0}^{+1.5}$ & -2.03$_{-0.20}^{+0.21}$ & -21.03$_{-0.50}^{+0.37}$ \vspace{4pt}\\
		 & Finkelstein & 2.2$^{+1.7}_{-1.1}$ & -2.00$^{+0.19}_{-0.18}$ & -20.91$^{+0.32}_{-0.39}$ \vspace{4pt}\\
        \hline
        8 & & & &\vspace{4pt}\\
		 & \citet{Bouwens2021} & 0.9$_{-0.5}^{+0.9}$ & -2.23$_{-0.20}^{+0.20}$ & -20.93$_{-0.28}^{+0.28}$ \vspace{4pt}\\
		 & Bouwens & 1.9$^{+1.3}_{-0.9}$ & -1.93$^{+0.17}_{-0.16}$ & -20.72$^{+0.28}_{-0.33}$ \vspace{4pt}\\
		 & \citet{Finkelstein2015} & 0.7$_{-0.7}^{+2.5}$ & -2.36$_{-0.40}^{+0.54}$ & -20.89$_{-1.08}^{+0.74}$ \vspace{4pt}\\
		 & Finkelstein & 2.8$^{+4.2}_{-2.0}$ & -2.20$^{+0.44}_{-0.38}$ & -20.32$^{+0.45}_{-0.56}$ \vspace{4pt}\\
        \hline
	\end{tabular}
\end{table*}

\subsection{Comparison of luminosity function parameters}\label{hst_sec:LFresultscomparison}

Figure~\ref{hst_fig:LFMeta} compares the Schechter function parameter measurements using our method with the results of \cite{Finkelstein2015} and \citet{Bouwens2021}. While our results agree broadly with these works, we do differ in the details.
This is not surprising, as our method is more constrained when it comes to the normalization of each individual field and allows for slightly different luminosity function shapes through cosmic variance \citep{Trapp2020,Trapp2022}. 
In particular, compared to \citet{Bouwens2021}, our framework prefers a 1-$\sigma$ lower $\phi^*$ and 1.5-$\sigma$ lower $\mabs^*$ at $z =$ 6. At $z =$ 7, our framework prefers a 1-$\sigma$ higher $\phi^*$ and 1.5-$\sigma$ shallower $\alpha$. At $z = 8$, our framework prefers a 1-$\sigma$ higher $\phi^*$ and 1-$\sigma$ shallower $\alpha$. In their work, \citet{Bouwens2021} include a treatment of the uncertainty in measured source luminosity, an effect not considered in this work. This would have the strongest effect at the bright end of the luminosity function, and could contribute to the different findings for the highly-correlated $\mabs^*$ and $\phi^*$.  
Despite these differences, our best-fit luminosity function matches the total number density of sources measured by \citet{Bouwens2021} between $m_{\textrm{app}} = 26-29$ within 10\%  at all redshifts.
Our framework recovers the  \citet{Finkelstein2015} results within 1-$\sigma$ across the board. However, our best-fit luminosity function predicts 10\%, 20\%, and 35\% more sources than \citet{Finkelstein2015}'s at $z =$ 6, 7, and 8, respectively (between $m_{\textrm{app}} = 26-29$).

\begin{figure*}\label{hst_fig:realEnvs}
    \centering
    \includegraphics[width=\textwidth]{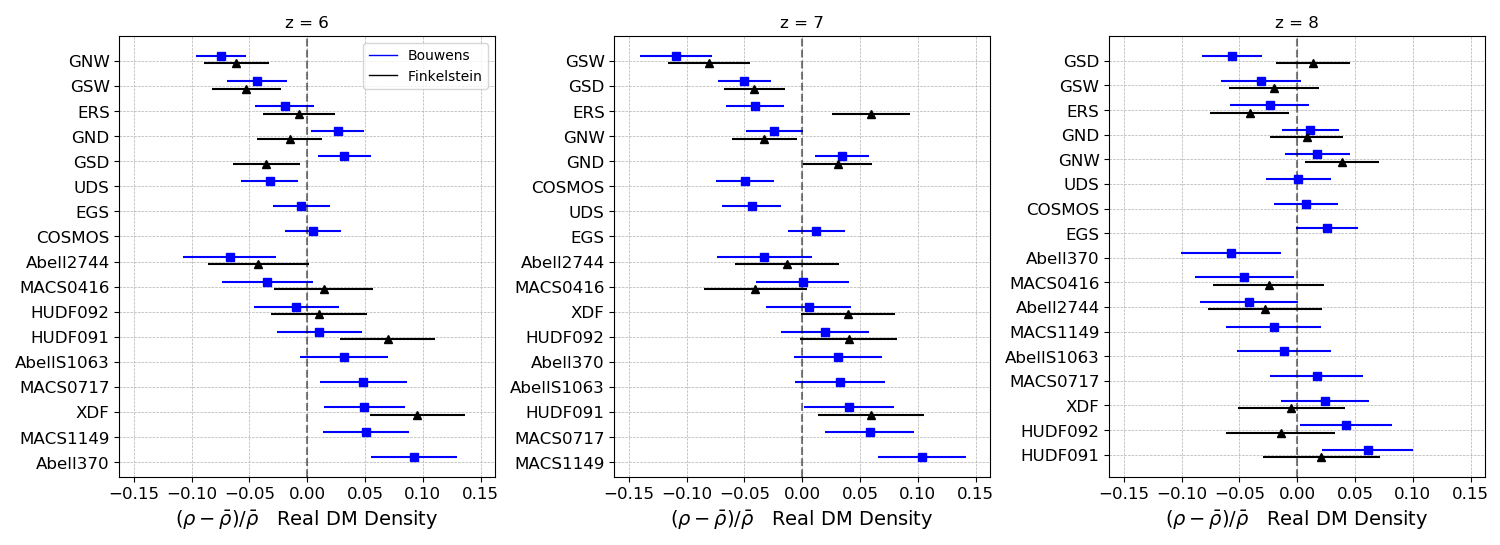}
    \caption{The relative over-densities of each field and their 68.27\% credible intervals.}
\end{figure*}

\subsection{Exploring systematics}\label{hst_sec:systematics}

There are many subtleties that affect the measurement of the luminosity function. 
For example, in our framework, the choice of where to cut off the faintest sources when fitting the luminosity function can have a $\sim$1-$\sigma$ effect on the resulting parameters, especially for the faint-end slope. This systematic will be immediately improved in the first cycle of JWST data, with multiple large and deep galaxy surveys like PRIMER, CEERS, JADES, PANORAMIC, WDEEP, and COSMOS-Web. At present, differing treatments of the faint end contribute to the differences between methods shown here. 

Different groups also have different selection criteria and analysis pipelines. They may therefore be probing slightly different populations of galaxies or physical locations.
For example, the results reported in \citet{Bouwens2021} and \citet{Finkelstein2015} agree with one another at the level of 1-2-$\sigma$, with \citet{Finkelstein2015} preferring a lower normalization parameter and steeper faint-end slope. 
At $z=6$, 7, and 8, the best fit luminosity functions from \citet{Finkelstein2015} predict 45\%, 32\%, and 
3\% fewer sources than the best fit luminosity functions from \citet{Bouwens2021} in the range $m_{\textrm{app}} = 26 - 29$.
Similarly, at $z=6$, 7, and 8, the best fit luminosity function from \citet{Finkelstein2015} predicts 37\% less, 36\% less, and 45\% \textit{more} total star formation rate density than the best fit luminosity function from \citet{Bouwens2021} (integrating down to $M_{\textrm{abs}} = -13$).
The exact reasons for these differences are not clear, and may be a combination of one or more of the following:
\textit{(i)} differences in methodology for generating effective volume curves \citep[which is most important at the faint-end, see][for details about these methodologies]{Finkelstein2015,Bouwens2021}; \textit{(ii)} differences in redshift intervals being probed, leading to sampling different physical volumes (see Table~\ref{hst_tab:fields}); 
\textit{(iii)} differences in the intrinsic galaxy population arising from the detailed selection criteria; 
and \textit{(iv)} other systematics.
In the next section, we analyze the individual environments of each field using both groups' data sets. Discrepancies in these densities could help illuminate differences between the groups.

In Figure~\ref{hst_fig:LFs}, the \citet{Bowler2014,Bowler2015,Bowler2020} best-fit binned luminosity function points (\textit{open squares}) lay somewhat below the best-fit lines at $z=$ 6 and especially at $z=$ 7 when compared to the \citet{Bouwens2021} data set (\textit{left column}). However, the points agree more closely with \citet{Finkelstein2015} data set (\textit{right column}). This highlights the potential effects of systematics inherent in different reduction techniques as well as space- vs ground-based measurements. However, there is considerable Poisson variance in the space-based data at these bright magnitudes due to low numbers of sources, which could account for these differences. Future wide-field space-based surveys such as those conducted by the Roman Space Telescope will be very important in measuring this bright-end cutoff.

Some studies of the high-$z$ galaxy distribution have found evidence that a Schechter function does not provide a good fit to the bright end of the luminosity function \citep[e.g.,][]{Bowler2014,Bowler2015,Bowler2020}.
To test this, we fit the data again using a modified Schechter function where we change the exponential factor to $e^{(L/L^*)^\varGamma}$, with $\varGamma$ a constant parameter.
When $\varGamma$ is less than one, this has the effect of flattening out the exponential cutoff, resulting in more bright galaxies. We find adding this extra parameter is disfavored given the data-set we use, increasing the Bayesian Information Criteria (BIC) by $\sim$5.
JWST alone will provide strong constraints on the shape of the luminosity function at the bright-end, especially at $z \sim 6$. At higher redshifts, the Roman Telescope will be crucial in measuring the bright-end of the luminosity function, especially because its data will be easily comparable to JWST due to the considerable overlap in their observable magnitudes.
Wide-field ground-based surveys
will also help in measuring the brightest galaxies. However, these ground-based surveys are limited in the redshifts they can reach, and when combining deep but narrow space-based images with shallow but wide ground-based images, one must consider carefully potential systematic normalization offsets between space- and ground-based measurements.

\section{Measurements of Large-Scale Structure}\label{hst_sec:Envresults}

\subsection{Survey Field Environments}\label{hst_sec:fieldenvs}

Figure~\ref{hst_fig:realEnvs} shows the physical dark matter densities of each Hubble field $\rho$ relative to the average dark matter density of the Universe $\bar{\rho}$. We list the numerical results in Table~\ref{hst_tab:densities}. We display the results when using the \citet{Bouwens2021} data-set as well as when using the \citet{Finkelstein2015} data-set. In general, the results are consistent between groups, with a few exceptions that will be discussed below.
The \textit{normalized} relative densities (in units of standard deviations $\sigmapb$ from average) are also given in Table~\ref{hst_tab:densities}.

We convert from a dark matter over/under-density to a theoretical $M^*$ galaxy number over/under-density using the bias value calculated using our public Python package \pakidge~\citep{Trapp2020}.
Typical bias values for an $M^*$ galaxy these surveys are 6--9.
At $z =$ 6, the densest \citet{Bouwens2021} field is Abell 370, with 9\% more dark matter than the Universe average for that volume, corresponding to a 60\% overdensity in $M^*$ galaxies. The least dense fields are Abell 2744 and GOODS-North Wide, with 7\%/7\% less matter than average and 43\%/48\% fewer $M^*$ galaxies. At $z =$ 7, the densest field is MACS1149 with 10\% more matter and 80\% more $M^*$ galaxies than average, and the least dense field is GOODS-South Wide with 11\% less matter and 90\% fewer $M^*$ galaxies than average. Finally, at $z =$ 8, the densest field is HUDF091 with 6\% more matter and 45\% more $M^*$ galaxies than average, and the least dense field is Abell 370 with 6\% less matter and 47\% fewer $M^*$ galaxies than average. At $z =$ 8 however, the uncertainty is much higher in these measurements and there is more disagreement between the data-sets.

\begin{table*}
	\centering
	\caption{The \textit{real} field densities, $\delta = (\rho - \bar{\rho})/\bar{\rho}$, and the  \textit{normalized} field densities, $\delta/\sigma_{\textrm{PB}}$, 
	with their uncertainties.}
	\label{hst_tab:densities}
	\begin{tabular}{rrrrrrr}
		\hline
		\hline
		\multicolumn{1}{c}{Field \& Data-set} & \multicolumn{1}{c}{z = 6} & & \multicolumn{1}{c}{z = 7} & & \multicolumn{1}{c}{z = 8} \\
		\multicolumn{1}{c}{$[B]$ouwens or $[F]$inkelstein} & \multicolumn{1}{c}{[norm]} & \multicolumn{1}{c}{[real]} & \multicolumn{1}{c}{[norm]} & \multicolumn{1}{c}{[real]} & \multicolumn{1}{c}{[norm]} & \multicolumn{1}{c}{[real]}\\
		\hline
		 & & & & & &\\
		CANDELS-GS-DEEP $[B]$ & 0.8$\pm$0.5 & 0.03$\pm$0.02 & -1.2$\pm$0.6 & -0.05$\pm$0.02 & -1.5$\pm$0.7 & -0.06$\pm$0.03\\
		CANDELS-GS-DEEP $[F]$ & -0.8$\pm$0.7 & -0.04$\pm$0.03 & -1.0$\pm$0.7 & -0.04$\pm$0.03 & 0.4$\pm$0.8 & 0.01$\pm$0.03\vspace{4pt}\\
		ERS $[B]$ & -0.4$\pm$0.6 & -0.02$\pm$0.03 & -0.9$\pm$0.6 & -0.04$\pm$0.02 & -0.6$\pm$0.8 & -0.02$\pm$0.03\\
		ERS $[F]$ &-0.2$\pm$0.7 & -0.01$\pm$0.03 & 1.4$\pm$0.8 & 0.06$\pm$0.03 & -1.0$\pm$0.8 & -0.04$\pm$0.03\vspace{4pt}\\
		CANDELS-GS-WIDE $[B]$ & -0.9$\pm$0.6 & -0.04$\pm$0.03 & -2.5$\pm$0.7 & -0.11$\pm$0.03 & -0.7$\pm$0.8 & -0.03$\pm$0.03\\
		CANDELS-GS-WIDE $[F]$ & -1.1$\pm$0.6 & -0.05$\pm$0.03 & -1.8$\pm$0.8 & -0.08$\pm$0.04 & -0.5$\pm$0.9 & -0.02$\pm$0.04\vspace{4pt}\\
		CANDELS-GN-DEEP $[B]$ & 0.6$\pm$0.5 & 0.03$\pm$0.02 & 0.9$\pm$0.6 & 0.03$\pm$0.02 & 0.3$\pm$0.7 & 0.01$\pm$0.02\\
		CANDELS-GN-DEEP $[F]$ & -0.4$\pm$0.7 & -0.02$\pm$0.03 & 0.8$\pm$0.8 & 0.03$\pm$0.03 & 0.2$\pm$0.8 & 0.01$\pm$0.03\vspace{4pt}\\
		CANDELS-GN-WIDE $[B]$ & -1.8$\pm$0.5 & -0.07$\pm$0.02 & -0.6$\pm$0.6 & -0.02$\pm$0.02 & 0.5$\pm$0.8 & 0.02$\pm$0.03\\
		CANDELS-GN-WIDE $[F]$ & -1.5$\pm$0.7 & -0.06$\pm$0.03 & -0.8$\pm$0.7 & -0.03$\pm$0.03 & 1.0$\pm$0.9 & 0.04$\pm$0.03\vspace{4pt}\\
		HUDF/XDF $[B]$ & 0.8$\pm$0.6 & 0.05$\pm$0.03 & 0.1$\pm$0.7 & 0.01$\pm$0.04 & 0.5$\pm$0.7 & 0.02$\pm$0.04\\
		HUDF/XDF $[F]$ & 1.6$\pm$0.7 & 0.10$\pm$0.04 & 0.7$\pm$0.7 & 0.04$\pm$0.04 & -0.1$\pm$0.9 & -0.00$\pm$0.05\vspace{4pt}\\
	    HUDF09-1 $[B]$ & 0.2$\pm$0.6 & 0.01$\pm$0.04 & 0.7$\pm$0.7 & 0.04$\pm$0.04 & 1.1$\pm$0.7 & 0.06$\pm$0.04\\
		HUDF09-1 $[F]$ & 1.2$\pm$0.7 & 0.07$\pm$0.04 & 1.1$\pm$0.8 & 0.06$\pm$0.05 & 0.4$\pm$0.9 & 0.02$\pm$0.05\vspace{4pt}\\
		HUDF09-2 $[B]$ & -0.2$\pm$0.6 & -0.01$\pm$0.04 & 0.4$\pm$0.7 & 0.02$\pm$0.04 & 0.8$\pm$0.7 & 0.04$\pm$0.04\\
		HUDF09-2 $[F]$ & 0.2$\pm$0.7 & 0.01$\pm$0.04 & 0.7$\pm$0.7 & 0.04$\pm$0.04 & -0.3$\pm$0.9 & -0.01$\pm$0.05\vspace{4pt}\\
		MACS0416-Par $[B]$ & -0.6$\pm$0.7 & -0.03$\pm$0.04 & 0.0$\pm$0.7 & 0.00$\pm$0.04 & -0.8$\pm$0.8 & -0.05$\pm$0.04\\
		MACS0416-Par $[F]$ & 0.2$\pm$0.7 & 0.01$\pm$0.04 & -0.7$\pm$0.8 & -0.04$\pm$0.04 & -0.5$\pm$0.9 & -0.02$\pm$0.05\vspace{4pt}\\
		Abell 2744-Par $[B]$ & -1.1$\pm$0.7 & -0.07$\pm$0.04 & -0.6$\pm$0.7 & -0.03$\pm$0.04 & -0.8$\pm$0.8 & -0.04$\pm$0.04\\
		Abell 2744-Par $[F]$ & -0.7$\pm$0.7 & -0.04$\pm$0.04 & -0.2$\pm$0.8 & -0.01$\pm$0.04 & -0.5$\pm$0.9 & -0.03$\pm$0.05\vspace{4pt}\\
		\hline
		CANDELS-UDS $[B]$ & -0.9$\pm$0.7 & -0.03$\pm$0.02 & -1.3$\pm$0.7 & -0.04$\pm$0.03 & 0.0$\pm$0.9 & 0.00$\pm$0.03\vspace{4pt}\\
		CANDELS-COSMOS $[B]$ & 0.1$\pm$0.7 & 0.01$\pm$0.02 & -1.5$\pm$0.7 & -0.05$\pm$0.02 & 0.2$\pm$0.9 & 0.01$\pm$0.03\vspace{4pt}\\
		CANDELS-EGS $[B]$ & -0.1$\pm$0.7 & -0.01$\pm$0.02 & 0.4$\pm$0.7 & 0.01$\pm$0.02 & 0.8$\pm$0.8 & 0.03$\pm$0.03\vspace{4pt}\\
		MACS0717-Par $[B]$ & 0.8$\pm$0.6 & 0.05$\pm$0.04 & 1.0$\pm$0.7 & 0.06$\pm$0.04 & 0.3$\pm$0.7 & 0.02$\pm$0.04\vspace{4pt}\\
		MACS1149-Par $[B]$ & 0.9$\pm$0.6 & 0.05$\pm$0.04 & 1.9$\pm$0.7 & 0.10$\pm$0.04 & -0.4$\pm$0.8 & -0.02$\pm$0.04\vspace{4pt}\\
		Abell S1063-Par $[B]$ & 0.6$\pm$0.6 & 0.03$\pm$0.04 & 0.6$\pm$0.7 & 0.03$\pm$0.04 & -0.2$\pm$0.8 & -0.01$\pm$0.04\vspace{4pt}\\
		Abell 370-Par $[B]$ & 1.6$\pm$0.6 & 0.09$\pm$0.04 & 0.6$\pm$0.7 & 0.03$\pm$0.04 & -1.1$\pm$0.8 & -0.06$\pm$0.04\vspace{4pt}\\
        \hline
	\end{tabular}
\end{table*}

We find, in general, our measurements of the dark matter density are consistent when using data from \citet{Finkelstein2015} or \citet{Bouwens2021}. However, GSD at $z =$ 6 and 8, and ERS at $z =$ 7, have very discrepant density measurements between the two analyses. The exact reasons for these differences are unknown; the potential culprits are likely the same as those for the differences in the determination of the luminosity function discussed in section~\ref{hst_sec:systematics}.

The XDF field lays inside the larger GOODS-S field, yet at $z =$ 6/7, their densities differ by more than one standard deviation (see Figure~\ref{hst_fig:realEnvs}). This is to be expected, as they are probing different scales. Using the excursion-set formalism \citep{Bond1991,Lacey1993}, we can calculate the expected root variance in the density field around a single point between two scales. Between scales defined by a 5 arcmin$^2$ (XDF) and 100 arcmin$^2$ (GOODS-S) with $z =$ 5.5 - 6.5, the root variance in the density field is 0.044. This variance combined with the uncertainties in the density measurements themselves makes it unsurprising to find e.g., an over-dense XDF field inside an under-dense GOODS-S field.

\begin{figure}
    \centering
    \includegraphics[width=0.5\textwidth]{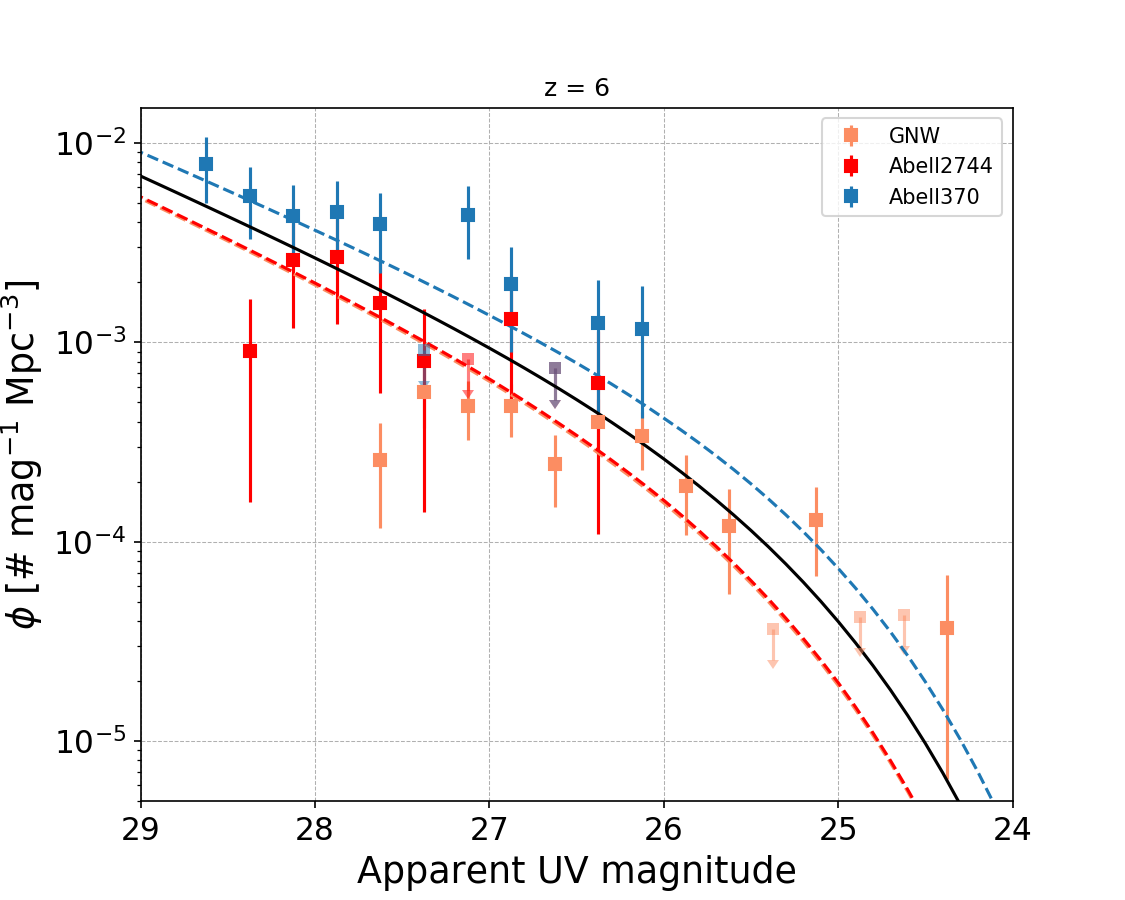}
    \caption{Data and best-fit luminosity functions $\Phi_{\textrm{obs}}$ for the three fields with the most extreme environments at $z =$ 6 in the \citet{Bouwens2021} data set. The \textit{black} curve is the best-fit average luminosity function. Abell370 and Abell2744 are clearly offset in normalization, but have very similar shapes in the magnitudes they cover. GNW is under-dense (\textit{orange-dashed} curve lies directly beneath \textit{red-dashed} curve), but its data points differ significantly from its best-fit luminosity function due to Poisson noise on the bright and faint ends.}
    \label{hst_fig:LFextreme}
\end{figure}

Finally, we find the most and least-dense fields have luminosity functions that are distinguishable by their normalization, but not by their shape (see Fig.~\ref{hst_fig:LFextreme}). JWST will be able to measure individual fields to much higher precision, potentially being able to distinguish over/under-densities by the shapes of their luminosity functions.

\subsection{Validation} \label{hst_sec:validation}

\subsubsection{Combining surveys}
In section~\ref{hst_sec:data}, we described creating composite surveys that cut down on the time it takes to fit the luminosity function. We tested the effect of combining surveys in this way by using an alternate grouping: folding the HUDF/XDF field into the \textbf{PAR} composite group. At all redshifts, this re-grouping does not affect the width of the resulting posterior, but it does shift its position in the following way. At $z =$ 6, the parameters $\phi^*$, $\alpha$, and $M^*$ decrease by 0.10$\sigma$, 0.18$\sigma$, and 0.15$\sigma$. At $z =$ 7, the same parameters decrease by 0.51$\sigma$, 0.76$\sigma$, and 0.48$\sigma$, respectively. Finally, at $z =$ 8, the same parameters \textit{increase} by 0.25$\sigma$, 0.47$\sigma$, and 0.30$\sigma$, respectively.
This difference most likely occurs because the XDF is significantly deeper than the other \textbf{PAR} fields, so its completeness function is reasonably different from the other fields as well.
Fields should only combined into composite fields if they have similarly-shaped effective volume curves and cover similar magnitude ranges, as is the case for the rest of our composite fields.
Even so, the modest changes resulting from the different treatment of the XDF demonstrate that our results are robust to the details of the composite fields.

\subsubsection{Post-processed environments}
In section~\ref{hst_sec:postprocess}, we described a process to measure the densities of each individual field, marginalizing over the posterior of the luminosity function $p(\vec{\phi} | \vec{D})$. Technically, this ``double-counts'' the fields, as each field was used to create that posterior in the first place. We test this by comparing the density of XDF generated by the full fitting framework (not double-counted) with the density of XDF using the ``post-processing'' described in section~\ref{hst_sec:postprocess} (double-counting). The densities are the same within 0.02$\sigma$ at all redshifts, showing the process is robust despite the double-counting. As described previously, the XDF has a relatively strong effect on the determination of $p(\vec{\phi} | \vec{D})$, so the double-counting effect is even smaller for less-influential fields.

\subsubsection{Measuring cosmic variance}\label{hst_sec:measuringcosmicvariance}

With the density measurements from Section~\ref{hst_sec:fieldenvs} in hand, one can ask: does the variance in measured densities match the expectation from theory? Given the range of physical scales probed by the fields we consider, this can only be explored in a global sense. That is, we
can evaluate our model of cosmic variance with our results by calculating the standard deviation of all of the \textit{normalized} density measurements $(\delta/\sigma_{\textrm{PB}})$ weighted by the sizes of the error bars. By definition, the standard deviation of the normalized densities should equal unity. If our model for cosmic variance gave values that were globally too large, we would expect the standard deviation of our measured normalized densities to be less than one, or greater than one if our model for cosmic variance was too small. This provides a way to test whether our theoretical inputs (including the conditional mass function and galaxy bias model) are reasonable. 

The standard deviation of the measured normalized densities is 0.92$\pm$0.09.
Therefore, the data do not prefer globally larger or smaller cosmic variance model. Doing the same calculation but splitting out into the different redshifts gives us standard deviation values of 0.90$\pm$0.16, 1.06$\pm$0.19, and 0.74$\pm$0.13 for redshifts 6, 7, and 8, respectively. Redshifts 6 and 7 do not prefer larger or smaller values of cosmic variance, but redshift 8 prefers a 2-$\sigma$ smaller value for cosmic variance.

The largest uncertainty in the \citet{Trapp2020} model of cosmic variance is the uncertainty in the conditional halo mass function -- the cosmic variance of dark matter haloes themselves. In \citet{Trapp2020}, we estimate this uncertainty to be $\sim$25\% globally. To test the effects of this uncertainty on our results, we allow the overall normalization of $\epcv$ to vary as a free parameter in one of our medium-resolution fits, constrained by a gaussian prior with relative width 0.25. We find that a globally larger/smaller value of $\epcv$ is not preferred, and marginalizing over this free parameter does not change the luminosity function posterior significantly. Therefore, to save computation time in the final, higher-resolution fits, we do not consider the uncertainty in $\epcv$. Such improvements will be important for deeper surveys with smaller errors on the density field.  

\subsection{Ionization Environment} \label{hst_sec:ionization}

In this section, we convert the density measurements from section~\ref{hst_sec:fieldenvs} to ionization states using a very simple mapping. This section serves as an example of how to make use of information on the large-scale density of a region. To do this, we construct a simple toy model of reionization. Our prescription can be made more rigorous by comparing to more detailed reionization models, such as those generated by \texttt{21cmFAST} \citep{Mesinger2011, Park2019}, but we focus here on a very simple prescription to make the inference as transparent as possible.

In this model, we assume the ionized fraction of hydrogen $Q$ in a region is linearly dependent on the fraction of baryons that have collapsed into haloes $f_{\textrm{coll}}$ through an efficiency parameter $\zeta$:
\begin{equation}\label{lae_eq:Qzetafcoll}
    Q = \zeta \cdot f_{\textrm{coll}}.
\end{equation}
 Within the Press-Schechter model \citep{Press1974, Lacey1993},
\begin{equation}\label{hst_eq:fcoll}
    f_{\textrm{coll}}(\delta, R_\alpha, z) = {\textrm{erfc}}\left( \frac{\delta_{\textrm{crit}}(z) - \delta_0}{\sqrt{2(\sigma^2_{\textrm{min}} - \sigma^2_{R})}} \right),
\end{equation}
where $\delta_{\textrm{crit}}(z)$ is the linearized density required for spherical collapse \citep[approximately 1.69 divided by the growth factor,][]{Eisenstein1998}, $\delta_0$ is the density of the region $\delta$ scaled to $z = 0$ (via the growth factor), $\sigma_{R}$ is the linear r.m.s fluctuation of the dark matter density field on a scale of $R$, and $\sigma_{\textrm{min}}$ is the same but on the scale of the smallest virialized halo that can form a galaxy, both evaluated at $z=$ 0. We take that smallest halo to have a virial temperature $T_{vir}=10^4K$, when atomic line cooling becomes efficient enough to collapse and fragment gas clouds  for star formation \citep{Loeb2013}. For concreteness, we then assert that reionization is complete for an \emph{average} region at $z=$ 6 (i.e., $Q = 1$ for a large region with $\delta_0=0$). This allows us to define $\zeta = 1 / f_\textrm{coll,6}$, where $f_\textrm{coll,6} = f_\textrm{coll}(z=6, \sigma_R = 0, \delta_0 = 0)$, the average collapse fraction of the Universe at $z=$ 6. We can then calculate the ionization fraction of each Hubble field by multiplying its collapse fraction (eq.~\ref{hst_eq:fcoll}) by $\zeta$, with the density of the region taken from Table~\ref{hst_tab:densities}. 

This simple model makes three major assumptions about reionization. First, we assume that it completes (for an average region) at $z=6$. This is simply to fix numbers; if reionization ended at a slightly different time, our general conclusions would not change. Second, it assumes that galaxies' ionizing efficiencies are independent of their masses. This is very unlikely to be true of actual galaxies (e.g., \citealt{Trenti2010, Tacchella2013, Mason2015, Behroozi2015, Furlanetto2017}), but for the relative ionized fractions across survey fields (the only purpose we require), it should be reasonably accurate. Third, it assumes that all ionizing photons produced in a given volume contribute to the ionization state of that volume, rather than being absorbed by Lyman-limit systems or escaping to help ionize other regions. The latter is likely a reasonable assumption, because recent measurements show that the mean free path is quite small at $z \sim 6$ \citep{Becker2021}. However, ignoring absorption by dense regions of the IGM will overestimate the actual ionized fraction, especially at the tail end of of reionization. For that reason, values $Q > 1$ should be interpreted as the production of an excess of ionizing photons. 

\begin{figure*}
    \centering
    \includegraphics[width=\textwidth]{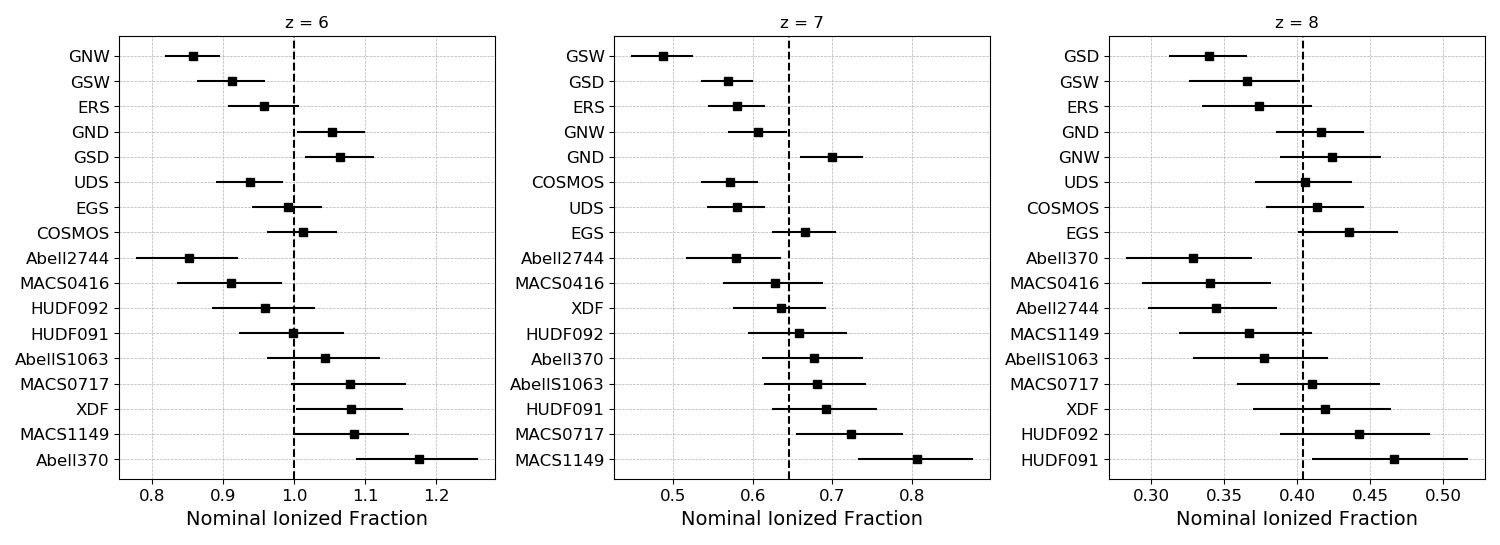}
    \caption{The nominal ionization state of the Hubble fields given a toy model of reionization and the density measurements from the \citet{Bouwens2021} data set (see Figure~\ref{hst_fig:realEnvs} and Table~\ref{hst_tab:densities}). The \textit{vertical dashed} lines correspond to the overall ionization of the Universe according to the model. As expected, the ionization states of the individual fields are distributed about the average ionization state of the Universe depending on their large-scale densities. The fields with  $Q>1$ can be interpreted as having reionized before the rest of the Universe. }
    \label{hst_fig:Qs}
\end{figure*}

We plot the results in Figure~\ref{hst_fig:Qs}.
The \textit{vertical dashed} lines in each panel correspond to the overall ionization of the Universe according to the model at each redshift. The fields with  $Q > 1$ can be interpreted as having reionized before the rest of the Universe. For example, the most dense field at $z=$ 6, Abell370, reionized at $z=6.4 \pm 0.2$ according to this model.

While these estimates of the ionization environments are obviously extremely crude, they demonstrate that even on the large scales of some HST surveys, the density fluctuations are large enough to cause substantial differences in the progress of reionization between survey volumes. For example, our results suggest that the $z=6$ GOODS North Wide field is delayed in its reionization compared to the GOODS South Wide field. Such inferences provide targets for investigations of the interplay between reionization and galaxy formation.



\section{Conclusions}\label{hst_sec:conclusions}

We measure the universal luminosity function of galaxies at $z =$ 6--8 using existing data \citet{Finkelstein2015,Bouwens2021}. We use a new fitting method that is more constrained than existing methods and also incorporates the density of individual fields or composited groups of fields \citep{Trapp2022}. Our results are consistent with existing studies \citep{Finkelstein2015,Bouwens2021}, but differ in the details (see Fig.~\ref{hst_fig:LFMeta}).
Our method has the benefit of considering the shape change of the luminosity function for different densities, an effect that will become more pronounced at higher redshift \citep{Trapp2020} and deeper observations.

We measure the dark matter density of most deep Hubble galaxy fields from $z =$ 6--8. We find the least/most-dense Hubble deep fields at $z =$ 6, 7, 8 are GNW/Abell370, GSW/MACS1149, and Abell370/HUDF091, respectively. These fields have expected dearths/excesses of $M^*$ galaxies of 
-48\%/60\%, -90\%/80\%, and -47\%/45\%, respectively.
We find dark matter densities are distributed in a way that is consistent with current estimations of cosmic variance \citep{Trapp2020,Bhowmick2020,Ucci2021}.
JWST will obtain many more dark-matter measurements of survey fields and at a higher precision than currently possible. These densities can be sorted and used to compare many statistical aspects of galaxies in under/over-dense environments, from the shape of the luminosity function, to the star-formation histories of galaxies, to the number of LAEs or QSOs in a region.
For example, in Figure~\ref{hst_fig:Qs}, we used a simple reionization model to associate the underlying density of the field with its ionization state, showing that even large-scale surveys (such as the GOODS fields) can have substantially different ionization states.

The pencil-beam shape of these volumes make interpreting a \textit{high} density complicated, as galaxies are likely clustered radially within the pencil-beam.
If galaxies can be sorted more precisely in redshift space (through accurate photometric redshifts, for example), it will be possible to make field density estimates on smaller, more spherical volumes, which will allow for closer comparison to observations.
In particular, our method could be improved to incorporate the probability distributions of photometric redshifts (and luminosities) as measurements improve.

We do not analyze the density sub-structure of fields, something that would be especially useful for large contiguous fields like COSMOS-Web. This is in-principle doable, as we will have 3-D positions of each source in each field.
We do not attempt this here because of the substantial radial widths of the HST fields, but it should be possible with JWST.



\section*{Acknowledgements}

We would like to thank our reviewer for their comments and suggestions.
We thank R. Bouwens and S.~L. Finkelstein for sharing their source catalogs and completeness functions for their surveys.
We also thank R.~Bowler for helpful discussions and R.~Bouwens for helpful comments on the manuscript.

This work was supported by the National Science Foundation through award AST-1812458. In addition, this work was directly supported by the NASA Solar System Exploration Research Virtual Institute cooperative agreement number 80ARC017M0006. We also acknowledge a NASA contract supporting the ``WFIRST Extragalactic Potential Observations (EXPO) Science Investigation Team'' (15-WFIRST15-0004), administered by GSFC. 

\textit{Software used:} This work uses iPython \citep{Perez2007} and the following Python packages: Matplotlib \citep{Hunter2007}, pandas \citep{McKinney2010}, NumPy \citep{Walt2011}, and SciPy \citep{Virtanen2020}.

\section*{Data Availability}

We include all luminosity function fits and density results in Tables~\ref{hst_tab:params} and \ref{hst_tab:densities}.
We provide the Schechter function posterior fits for both the \citet{Bouwens2021} and \citet{Finkelstein2015} data sets at $z =$6, 7, and 8 as supplementary data.




\bibliographystyle{mnras}
\bibliography{me} 





\bsp	
\label{lastpage}
\end{document}